\documentclass[a4paper,twocolumn,final]{revtex4}
\usepackage{amsmath}
\usepackage{mathtools}
\usepackage{amssymb}
\usepackage{color}
\usepackage{graphicx}
\usepackage{hyperref}
\usepackage{url}
\newcommand{\al}{\alpha}

\newcommand{\de}{\delta}
\newcommand{\De}{\Delta}
\newcommand{\eps}{\varepsilon}
\newcommand{\ep}{\varepsilon}
\newcommand{\fr}{\frac}

\newcommand{\la}{\langle}
\newcommand{\lf}{\left}

\newcommand{\om}{\omega}

\newcommand{\ra}{\rangle}
\newcommand{\rg}{\right}

\newcommand{\Si}{\Sigma}

\newcommand{\si}{\sigma}
\newcommand{\Te}{\Theta}
\newcommand{\te}{\theta}

\newcommand{\scr}{\mathcal}

\begin{document}
\title{Spectral and Parametric Averaging for Integrable Systems}
\author{Tao Ma}
\affiliation{Department of Physics, University of Cincinnati, Cincinnati, OH 45221-0011}
\author{R.A. Serota}
\email{serota@ucmail.uc.edu}
\affiliation{Department of Physics, University of Cincinnati, Cincinnati, OH 45221-0011}
\begin{abstract}
We analyze two theoretical approaches to ensemble averaging for integrable systems in quantum chaos - spectral averaging and parametric averaging. For spectral averaging, we introduce a new procedure - rescaled spectral averaging. Unlike traditional spectral averaging, it can describe the correlation function of spectral staircase and produce persistent oscillations of the interval level number variance. Parametric averaging, while not as accurate as rescaled spectral averaging for the correlation function of spectral staircase and interval level number variance, can also produce persistent oscillations of the global level number variance and better describes saturation level rigidity as a function of the running energy. Overall, it is the most reliable method for a wide range of statistics.
\end{abstract}

\maketitle

\section{Introduction}

The framework of quantum chaos is structured around the concept of ensemble averaging (EA). Statistics, such as correlation function of level density \cite{Wickramasinghe2005}, interval level number variance (IV) \cite{Wickramasinghe2005}, global level number variance (GV) \cite{ma12gv}, spectral rigidity (SR) \cite{Berry1985} and nearest neighbor spacing distribution \cite{ma10lr} are defined through EA. In the literature, two methods are employed to achieve EA for integrable systems. Traditionally, EA in semiclassical theories was understood in terms of spectral averaging (SA). \cite{Berry1985, Berry1986} A numerical simulation of IV using SA was performed in \cite{Grosche1999}. The oscillations of IV were found to decay, while the other EA method for integrable systems -- parametric averaging (PA) -- correctly showed persistent oscillations. \cite{Wickramasinghe2005} We explained that SA tends to suppress the non-decaying oscillatory behavior due to destructive interference of the running-energy-dependant non-coherent terms.\cite{Wickramasinghe2005}  Moreover, this paper will show that when, in order to avoid such destructive interference, SA is performed over a short range of sampled energies, sampling is insufficient and sample-specific fluctuations are observed.

Much as impurity averaging in disordered systems, PA corresponds to EA for a fixed value of the running energy; specifically for rectangular billiards (RB) averaging is over an ensemble of rectangles of varying aspect ratios and fixed area. To our knowledge, PA for integrable system was first performed by Casati \emph{et. al.} to prove the saturation of SR of integrable system. In their words, ``computed an average $\overline{\De}_3(L)$ ... by averaging $\De_3$ over a number of different values of $\al$ chosen at random in a given interval (`ensemble averaging').''\cite{Casati1985} ($\overline{\De}_3(L)$ denotes SR over the interval $L$ and $\al$ is the aspect ratio defined in \cite{Casati1985}.) PA with better implementations was used to reproduce saturation of SR \cite{Wickramasinghe2008, ma10lr}, produce persistent oscillations of IV \cite{Wickramasinghe2005, Wickramasinghe2008, ma10mk, ma11eb} and GV \cite{ma12gv} and prove level repulsion in integrable systems \cite{ma10lr}. These studies demonstrated that PA is a reliable and versatile method for numerical computation of statistics of integrable systems. (Note that PA can also be used as an experimental technique to study orbital magnetism. \cite{Levy1993}) 

Here we undertake a detailed comparison of SA and PA previously unaddressed in literature. The central result of this work is that, unlike PA, traditional SA cannot produce persistent oscillations of IV and GV. Even with our newly proposed \emph{rescaled spectral averaging} (RSA), one can only address IV oscillations. These results are argued theoretically and proved numerically.

This paper is organized as follows. In Sec. \ref{sec:statistics}, we review the periodic orbit (PO) theory of level fluctuations and the semiclassical theory of IV, GV, SR, and correlation function of spectral staircase (CFSS). In Sec. \ref{sec:theory_EA},  we define SA and PA for IV, CFSS, and SR. From linear expansion of SA, we argue that SA suppresses the oscillations of IV when averaging over large intervals. To preserve persistent oscillations, we propose RSA. In Sec. \ref{sec:num_simulation}, we present spectral fluctuations, IV, and GV computed from SA and PA, IV from RSA, CFSS from PA and RSA, and SR from SA, PA and RSA. In Sec. \ref{sec:summary}, we discuss advantages and shortcomings of RSA and PA and outline their applicability.

In this paper, we use uppercase letters to indicate ensemble averaged statistics and lowercase ones to indicate their corresponding sample-specific values. For instance, $\Si$ denotes IV and $\si$ denote sample IV; $\Si_g$ denotes GV and $\si_g$ denotes sample GV; $\De_3$ denotes SR and $\de_3$ denotes sample SR; $K_\scr{N}$ denotes CFSS and $k_\scr{N}$ denotes sample CFSS. The subscripts $A$ and $\Te$ indicate numerical computation and theoretical calculation respectively. The superscript indicates the EA method, that is SA, RSA or PA.

\section{Statistics}\label{sec:statistics}

\subsection{Periodic orbit theory of level fluctuations}

We use RB as a model system to illustrate our theory. For a particle of mass $m$ in a RB with sides $a, b$ and aspect ratio $\al \equiv a^2 / b^2$, the eigenenergy with quantum numbers $n_{1,2}$ is given by
\begin{equation}
\ep_{n_1, n_2} = \fr{\pi^2 \hbar^2}{2m} \lf( \fr{n_1^2}{a^2} + \fr{n_2^2}{b^2} \rg) .
\end{equation}
The spectral staircase (SS) is defined as
\begin{equation}
\scr{N}(\ep) \equiv \sum_{n_1, n_2} \te(\ep - \ep_{n_1, n_2} ) ,
\end{equation}
where $\te$ is unit step function.
According to Weyl's formula, the ensemble-averaged SS is given by \cite{Gutzwiller1990, Grosche1999, Wickramasinghe2005}
\begin{equation}\label{eq:WeylFormula}
\la \scr{N}(\ep) \ra
= \fr{\ep}{\De} - \fr{\scr{S}}{2\sqrt{\pi \scr{A}}} \lf( \fr{\ep}{\De} \rg)^{1/2} +\fr{1}{4}
\end{equation}
where $\De = 2\pi\hbar^2/m\scr{A}$; $\scr{A}$ and $\scr{S}$ are the RB area and perimeter respectively; and $\la \, \ra$ denotes EA. The second and third terms are usually called ``perimeter correction" and ``corner correction" respectively. In previous works \cite{ma10mk, ma10lr}, we only considered the perimeter correction when unfolding the spectrum. In the present paper, we account for both terms. After unfolding the spectrum by (\ref{eq:WeylFormula}), the mean level spacing becomes unity and \cite{Wickramasinghe2005}
\begin{equation}\label{eq:EA_N_ep}
\la\scr{N(\ep)}\ra = \ep ,
\end{equation}
which would be correct for a perfect EA method and is approximately correct for SA and PA as will be shown in Sec. \ref{sec:num:fluc}.

From the PO theory, the fluctuation of level density is given by
\begin{equation}\label{eq:derho_PO_theory}
\begin{split}
\de\rho(\ep)
&\equiv \rho(\ep)  - \la \rho(\ep) \ra \\
&=\fr{2}{\hbar^{\mu+1}} \sum_\mathbf{M} \de_\mathbf{M} A_\mathbf{M} (\ep) \cos\lf[\fr{S_\mathbf{M}(\ep)}{\hbar} - \fr{\pi}{4} \rg]
\end{split}
\end{equation}
and the fluctuation of SS by
\begin{equation}\label{eq:deN_PO_theory}
\begin{split}
\de \scr{N}(\ep)
&\equiv\scr{N}(\ep) - \la \scr{N}(\ep)\ra \\
&= \fr{2}{\hbar^{\mu}} \sum_\mathbf{M} \fr{\de_\mathbf{M} A_\mathbf{M}(\ep)}{T_\mathbf{M}(\ep)}  \sin\lf[ \fr{S_\mathbf{M}(\ep)}{\hbar} - \fr{\pi}{4} \rg] .
\end{split}
\end{equation}
Here $\mu=(\nu-1)/2$, $\nu$ is the dimensionality of phase space and the period, action, and  amplitude of PO-$\mathbf{M}$ are given respectively by \cite{Berry1985}
\begin{equation}\label{eq:RB_parameter}
\begin{split}
&T_\mathbf{M}(\ep) = [2m(M_1^2 a^2 + M_2^2 b^2)/\ep]^{1/2} \\
&S_\mathbf{M}(\ep) = 2\ep T_\mathbf{M} \\
&A_\mathbf{M}^2(\ep) = m^2 a^2 b^2 / \pi^3 \ep T_\mathbf{M} ,
\end{split}
\end{equation}
with $\mathbf{M} = (M_1,M_2)$ and non-negative $M_{1,2}$ as winding numbers. Above
\begin{equation}
\de_\mathbf{M}=
\begin{cases}
0    &M_1 = M_2 = 0 \\
1/2 &\text{if only one of } M_1, M_2 \text{ is zero} \\
1    &\text{otherwise} .
\end{cases}
\end{equation}
Compared with \cite{Berry1985}, in (\ref{eq:derho_PO_theory}) and (\ref{eq:deN_PO_theory}), we have an extra factor $- \pi/4$ from a quantum mechanical calculation \cite{ma10mk}. In Sec. \ref{sec:num:fluc}, we show that this factor matters.

\subsection{Interval and global level number variance}

IV is defined as
\begin{equation}\label{eq:IV_def}
\Si(\ep, E)
\equiv \la [ N - \la N \ra ]^2 \ra
= \la [ N - E ]^2 \ra  ,
\end{equation}
where $N \equiv \scr{N}(\ep_2) - \scr{N}(\ep_1)$ with
\begin{align}
\ep_1 &= \ep-E/2 \\
\ep_2 &= \ep+E/2
\end{align}
and $E\ll \ep$. GV is defined as \cite{Serota2008, ma12gv}
\begin{equation}\label{eq:GV_def}
\Si_g( \ep )
\equiv \la [ \scr{N}(\ep) - \la\scr{N}(\ep) ]^2 \ra
=  \la [ \scr{N}(\ep) - \ep ]^2 \ra  .
\end{equation}
In (\ref{eq:IV_def}) and (\ref{eq:GV_def}), we used (\ref{eq:EA_N_ep}).
We term
\begin{equation}\label{eq:sample_IV_def}
\si(\ep, E)
\equiv [ N - \la N \ra ]^2
=  [ N - E ]^2
\end{equation}
"sample IV" and
\begin{equation}\label{eq:sample_GV_def}
\si_g(\ep)
\equiv [ \scr{N}(\ep) - \la\scr{N}(\ep)\ra ]^2
= [ \scr{N}(\ep) - \ep ]^2
\end{equation}
``sample GV".

Employing the diagonal approximation (DA) \cite{Wickramasinghe2005, ma10mk}, theoretical sample IV is expressed as \cite{ma12gv}
\begin{equation}\label{eq:si_theory_not_unfolding_spectrum}
\si_\Te (\ep, E) = \fr{8}{\hbar^{2\mu}} \sum_\mathbf{M} \fr{\de_\mathbf{M}^2 A_\mathbf{M}^2(\ep)}{T_\mathbf{M}^2(\ep)}
\sin^2\lf( \fr{T_\mathbf{M}(\ep) E}{2\hbar} \rg) .
\end{equation}
Substituting (\ref{eq:RB_parameter}) and unfolding the spectrum, the above equation can be written as \cite{ma12gv}
\begin{equation}\label{eq:sample_IV_theory}
\si_\Te (\ep, E)=
4 \sqrt{\fr{\ep}{\pi^5 }}
\sum_{\mathbf M}
\fr{\de_\mathbf{M}^2}{R_\mathbf{M}^3} \sin^2\lf[ \sqrt{\fr{\pi}{\ep} } R_\mathbf{M} E \rg],
\end{equation}
where $R_\mathbf{M} = \sqrt{ M_1^2 \al^{1/2} + M_2^2 \al^{-1/2} }$.
Numerical sample IV, $\si_\text{A}(\ep, E)$, is a jagged line as a function of $E$, while theoretical sample IV, $\si_\Te(\ep, E)$, is a smooth line by (\ref{eq:sample_IV_theory}). EA is able to bridge this difference.

\subsection{Spectral rigidity}

SR is defined as
\begin{equation}\label{eq:SR_def}
\De_3 (\ep, E) \equiv
\lf\la \min_{(A,B)} \fr{1}{E} \int_{\ep_1}^{\ep_2} [\scr{N}(x) -A-Bx ]^2 dx \rg\ra ,
\end{equation}
which has the explicit form \cite{Berry1985}
\begin{equation}\label{eq:SR_explicit}
\begin{split}
&\Bigg\la \fr{1}{E} \int_{-E/2}^{E/2} \scr{N}^2(\ep+\om) d\om - \lf[\fr{1}{E} \int_{-E/2}^{E/2} \scr{N}(\ep+\om) d\om\rg]^2 \\
&-12\lf[ \fr{1}{E^2} \int_{-E/2}^{E/2} \om \scr{N}(\ep + \om) d\om \rg]^2 \Bigg\ra .
\end{split}
\end{equation}
Sample SR is defined as
\begin{equation}\label{eq:sample_SR_def}
\begin{split}
&\de_3 (\ep, E) \equiv
\min_{(A,B)} \fr{1}{E} \int_{\ep_1}^{\ep_2} [\scr{N}(x) -A-Bx ]^2 dx ,
\end{split}
\end{equation}
which is computed from (\ref{eq:SR_explicit}) without EA (that from the expression inside $\la \, \ra$).
The saturation SR $\De_3^\infty(\ep)$ and its sample value $\de_3^\infty(\ep)$ are numerically computed as $\De_3(\ep,E)$ and $\de_3(\ep,E)$ respectively with sufficiently large $E\gg \sqrt{\ep}$.
For the saturation SR, the ``minimization'' fit $A+B\ep$ is approximately given by $\ep$. Hence \cite{ma12gv}
\begin{equation}\label{eq:delta3_sample_GV}
\de_3 (\ep, E)\approx \fr{1}{E} \int_{\ep_1}^{\ep_2}  \si_g(x) dx ,
\end{equation}
where we used (\ref{eq:sample_GV_def}).

Based on DA, we have the sample value of saturation SR \cite{ma12gv}
\begin{equation}\label{eq:sample_delta3_theory}
(\de_3^\infty)_\Te (\ep) =   \sqrt{\fr{\ep}{\pi^5 }}
\sum_{\mathbf M}
\fr{\de_\mathbf{M}^2}{R_\mathbf{M}^3} ,
\end{equation}
where we used (\ref{eq:RB_parameter}) and unfolded the spectrum.

\subsection{Correlation function of spectral staircase}

CFSS is defined as \cite{Serota2008, ma12gv}
\begin{equation}\label{eq:KN_def}
K_\scr{N}(\ep, E) \equiv \la \de \scr{N}(\ep_1) \de \scr{N}(\ep_2) \ra .
\end{equation}
The sample CFSS is defined as
\begin{equation}\label{eq:Kdiag}
k_\scr{N}(\ep, E)
\equiv \de \scr{N}(\ep_1) \de \scr{N}(\ep_2) .
\end{equation}
Using DA, we have \cite{ma12gv}
\begin{equation}\label{eq:relation_kN_de3_si}
\begin{split}
k_\scr{N}(\ep, E)
&= \de_3^\infty(\ep) - \fr{1}{2}\si(\ep,E) \\
&\approx \de_3^\infty(\ep) - \fr{E}{2},  \qquad \text{for } E \ll \sqrt{\ep} .
\end{split}
\end{equation}
The ensemble averaged form is
\begin{equation}\label{eq:relation_KN_De3_Si}
K_\scr{N}(\ep, E) = \De_3^\infty(\ep) - \fr{1}{2}\Si(\ep,E) .
\end{equation}

\section{Theory of spectral and parametric Averaging}\label{sec:theory_EA}

\subsection{Spectral and parametric averaging}\label{subsec:defSAPA}

In SA, the numerical and theoretical values of IV can be respectively defined by the following integrals:
\begin{eqnarray}
\Si_{\text{A}}^\text{SA} (\ep, E) &\equiv
\int \si_\text{A}(x, E) f^\text{SA}(x) dx  \label{eq:IV_num_SA_def} \\
\Si_\Te^\text{SA} (\ep, E) &\equiv
\int \si_\Te(x, E) f^\text{SA}(x) dx , \label{eq:IV_theory_SA_def}
\end{eqnarray}
where $\si_\text{A}(x, E)$ and $\si_\Te(x, E)$ implicitly depend on the aspect ratio $\al_0$ and $f^\text{SA}(x)$ is the density of sampled energies and is chosen as equally spaced points in a range $\epsilon$ centered at $\ep$. In other words, $x \in [\eps-\epsilon/2, \eps+\epsilon/2]$ with uniform density.

In PA, the numerical and theoretical values of IV are respectively defined as
\begin{eqnarray}
&\Si_\text{A}^\text{PA} (\ep, E) \equiv
\int \si_\text{A} (\ep, E) f^\text{PA}(\al) d\al \label{eq:IV_num_PA_def}\\
&\Si_{\Te}^\text{PA} (\ep, E) \equiv
\si_\Te (\ep, E,\al_0), \label{eq:IV_theory_PA_al0_def}
\end{eqnarray}
where $\si_\text{A}(\ep, E)$ and $\si_\Te(\ep, E)$ implicitly depend on $\al$ and $f^\text{PA}(\al)$ is a Gaussian distribution with mean $\al_0$ and standard deviation $\ll 1$. Numerical computation of IV from SA and PA can be understood as numerical integration of (\ref{eq:IV_num_SA_def}) and (\ref{eq:IV_num_PA_def}) respectively. Similarly we can define SA and PA for CFSS and SR. \cite{PA}

\subsection{Linear expansion of spectral averaging}\label{subsec:linear_expansion_SA}

Using (\ref{eq:sample_IV_theory}), a representative term in (\ref{eq:IV_theory_SA_def}) reads
\begin{equation}\label{eq:SI_SA_term}
\begin{split}
&\int 4 \sqrt\fr{x}{\pi^5 }
\fr{\de_\mathbf{M}^2}{ R_\mathbf{M}^3} \sin^2\lf[\sqrt{\fr{\pi}{x} } R_\mathbf{M} E \rg] f^\text{SA}(x) dx \\
&\approx 4 \sqrt{\fr{\ep}{\pi^5 }}  \fr{\de_\mathbf{M}^2}{R_\mathbf{M}^3} \int_{\eps-\epsilon}^{\eps+\epsilon}
\sin^2\lf[\sqrt{\fr{\pi}{x} } R_\mathbf{M} E \rg] f^\text{SA}(x) dx .
\end{split}
\end{equation}
When $\epsilon$ is far larger than the period of the sine term, the integrand can be replaced by $1/2$ and one will not observe persistent oscillations of IV. The first-order derivative of the argument of sine is given by
\begin{equation}
\fr{d E \sqrt{\fr{\pi}{x}} R_\mathbf{M} }{dx} \bigg|_{x = \ep}
= -\fr{ \sqrt{\pi} R_\mathbf{M}  E }{2\ep^{3/2}} ,
\end{equation}
whereof we find that when
\begin{equation}\label{eq:SA_range}
\epsilon > \fr{2 \sqrt\pi \ep^{3/2}}{R_\mathbf{M} E}
\sim \fr{\ep^{3/2}}{E},
\end{equation}
oscillations of IV will decay. Notice that oscillations are observed when $E > \ep^{1/2}$ and that the decay of IV oscillations becomes faster with larger $E$. \cite{PA}

\subsection{Rescaled spectral averaging}

We just saw that traditional SA suffers from an inherent flaw due to destructive interference of oscillating terms. In order to observe persistent oscillations with larger and larger interval width $E$, one needs to sample sufficiently large energy range $\epsilon$ centered on $\ep$ to achieve proper EA. Yet, Eq. (\ref{eq:SA_range}) sets the limit to how large such range can be in order to avoid destructive interference and observe persistent oscillation of IV. Furthermore, the limit deceases with the increase of $E$. 

A possible workaround would be to sample various parts of spectrum, not necessarily around $\ep$. However, since persistent oscillations strongly depend on $\ep$ (the point of onset, the amplitude and the period \cite{Wickramasinghe2008, ma10mk}), such procedure, executed without a proper account for this $\ep$-dependence, would have an effect similar to the destructive interference above -- a wash-out of persistent oscillations. Consequently, we introduce a modified procedure, RSA, that allows sampling of different parts of the spectrum.

Our approach is based on a scaling transformation $\si_\Te(c\ep, \sqrt{c} E) = \sqrt{c} \si_\Te(\ep, E)$, which follows from (\ref{eq:sample_IV_theory}). Consequently, in RSA, when the running energy and the interval are scalled as $\ep \rightarrow c\ep$  and $E \rightarrow \sqrt{c} E$ respectively, $\si_\text{A}(c\ep,\sqrt{c} E)$ needs to be rescaled by a factor $1/\sqrt{c}$ before averaging. Numerically, IV is computed as
\begin{equation}
\Si_\text{A}^\text{RSA} (\ep, E) \equiv
\fr{1}{n+1} \sum_{i=0}^n \fr{1}{\sqrt{c_i}}
\si_\text{A} \lf(c_i\ep, \sqrt{c_i} E \rg) ,
\end{equation}
where $c_i$ is the ratio of the energy of a sampled spectral location to $\ep$ and $n+1$ is the number of sampled energies and theoretically, by design, it is given by 
\begin{equation}\label{eq:Si_RSA_theory}
\Si_\Te^\text{RSA} (\ep, E) = \si_\Te \lf(\ep, E \rg) ,
\end{equation}
that is coincides with (\ref{eq:IV_theory_PA_al0_def}). We note due to the close relation between IV and CFSS in (\ref{eq:relation_KN_De3_Si}), RSA can be similarly defined for the latter and we have
\begin{equation}\label{eq:KN_RSA_theory}
(K_\scr{N})_\Te^\text{RSA} (\ep, E) = (k_\scr{N})_\Te \lf(\ep, E \rg) ,
\end{equation}
where theoretical $(K_\scr{N})_\Te^\text{RSA} (\ep, E)$ can be evaluated from  (\ref{eq:sample_IV_theory}), (\ref{eq:sample_delta3_theory}), and (\ref{eq:relation_kN_de3_si}).

RSA of saturation SR is computed by
\begin{equation}
 \fr{1}{n+1}\sum_{i=0}^n \fr{1}{\sqrt{c_i}}  (\de_3^\infty)_\text{A} (c_i \ep)  ,
\end{equation}
and its theoretical value is
\begin{equation}\label{eq:saturation_SR_RSA_def}
(\De_3^\infty)_\Te^\text{RSA}(\ep) \equiv
(\de_3^\infty)_\Te (\ep).
\end{equation}
$(\De_3^\infty)_\text{A}^\text{RSA}(\ep)$ scales as $\sqrt{\ep}$ for billiard systems, including elliptic billiards \cite{ma11eb}.

\section{Numerical simulations}\label{sec:num_simulation}

Below, except in Fig. \ref{fig:deN_PA}, for SA and RSA, the aspect ratio $\al_0$ is set to be $1-(\sqrt{5}-1)/20 \approx 0.938$ to avoid degeneracy; for PA, the distribution of $\al$ is a Gaussian distribution with mean $\al_0$ and standard deviation 0.2. In the computation of IV and CFSS, we set $\ep=10^5$.\\

\subsection{Fluctuations of spectral staircase}\label{sec:num:fluc}

\begin{figure}
\begin{center}
\includegraphics[width=0.23\textwidth]{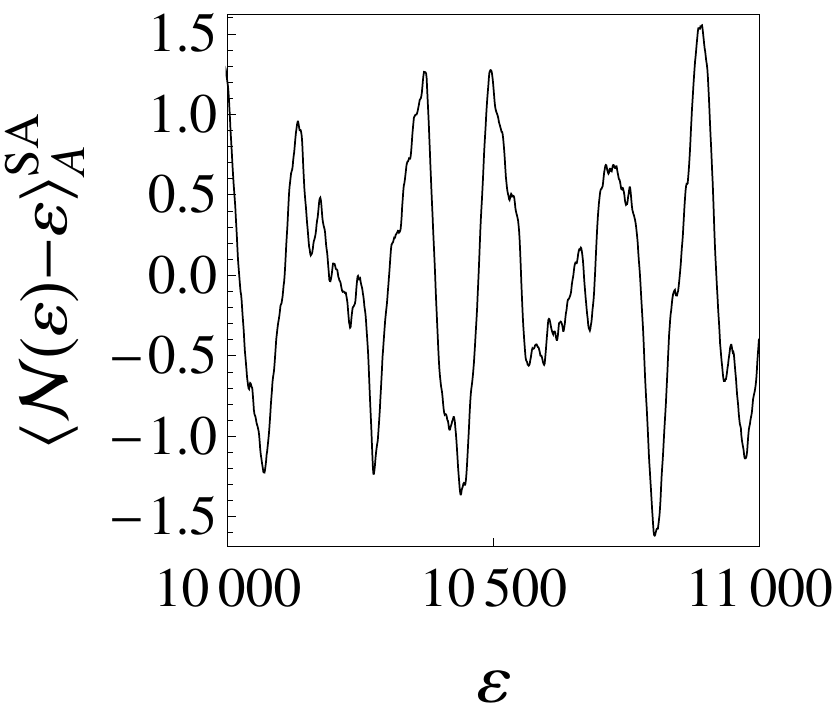}
\includegraphics[width=0.23\textwidth]{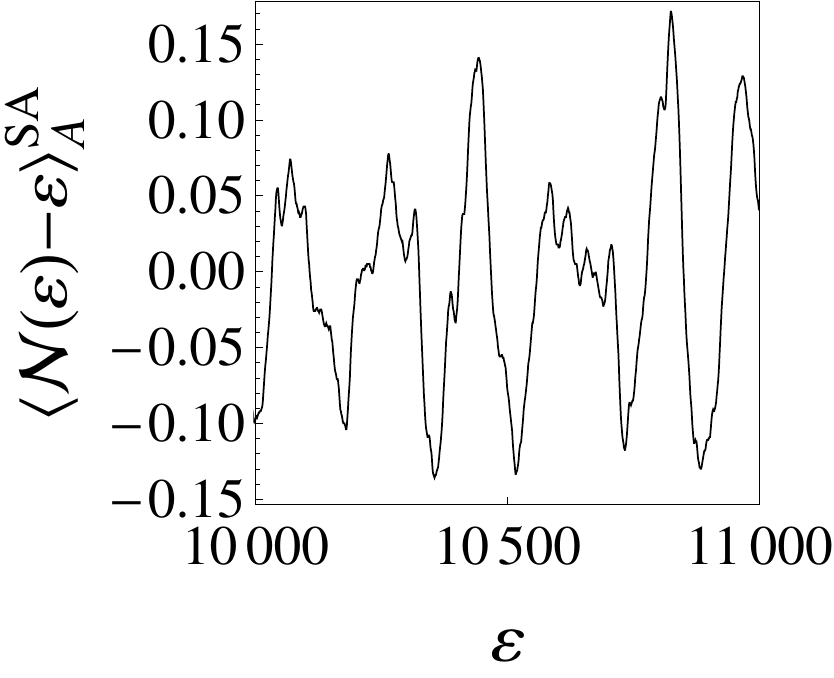}
\end{center}
\caption{Spectral averaging of the fluctuation of SS with different ranges of sampled energies. Left: the range is $10^2$; Right: $10^3$.}\label{fig:deN_SA}
\end{figure}

\begin{figure}
\begin{center}
\includegraphics[width=0.45\textwidth]{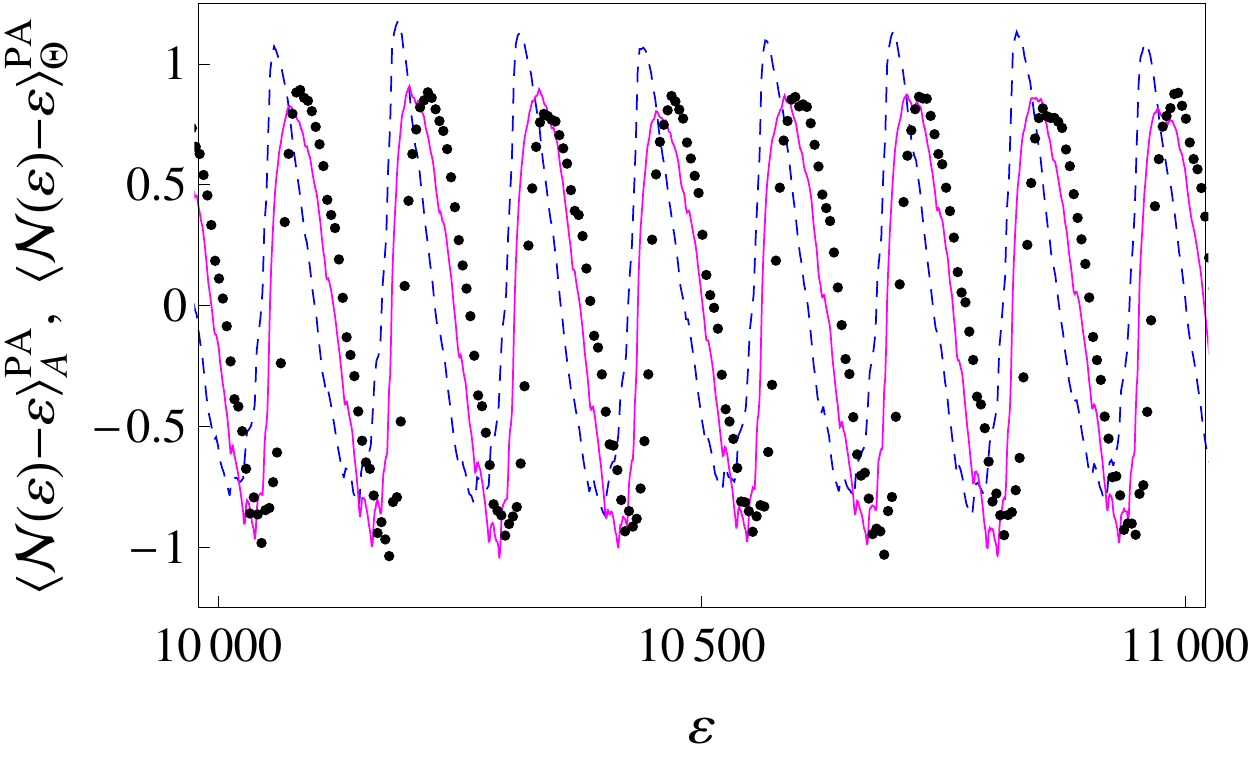}\\
\includegraphics[width=0.45\textwidth]{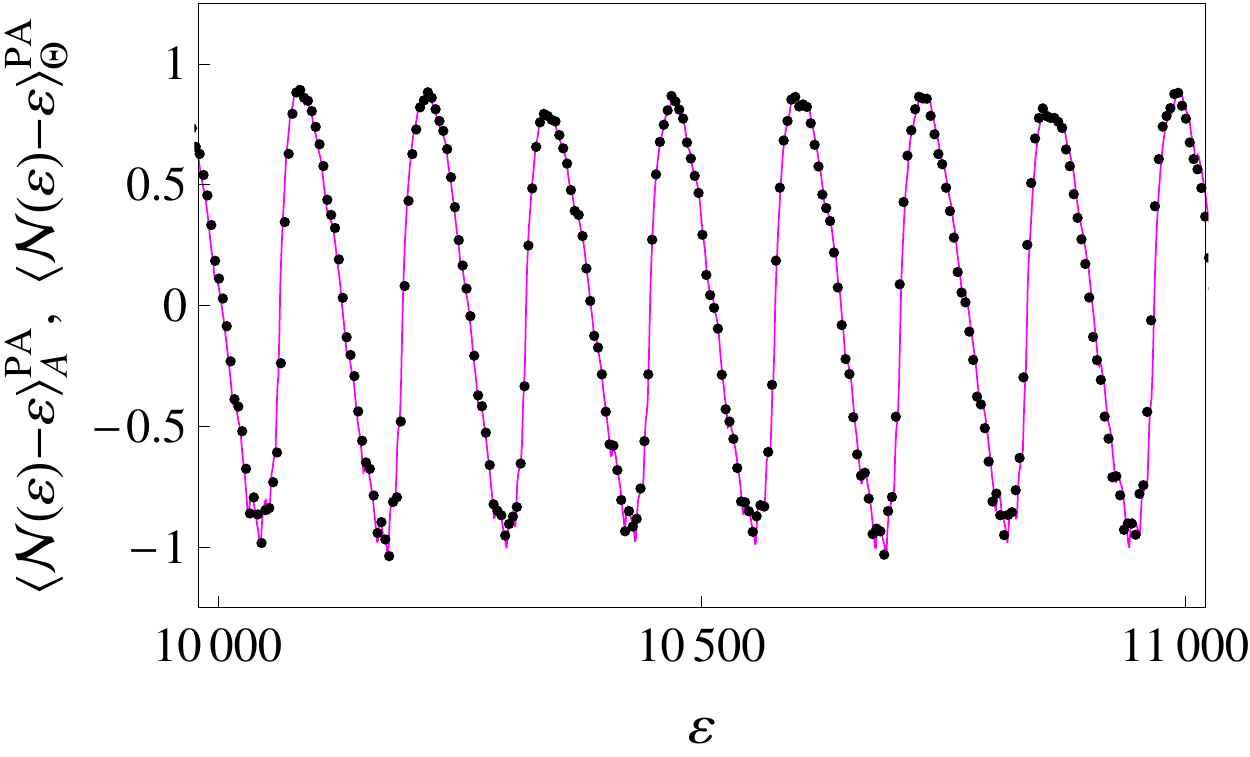}
\end{center}
\caption{Parametric averaging of the fluctuation of SS. Top: black dots, numerical $\la \scr{N}(\ep) - \ep \ra_\text{A}^\text{PA}$ calculated by averaging over $\al$ from a Gaussian distribution with the mean $1$ and the standard deviation 0.2; magenta line, $\la \scr{N}(\ep) - \ep \ra_\Te^\text{PA}$ calculated from (\ref{eq:deN_PO_theory}) and averaged over $\al$; dashed blue line, $\la \scr{N}(\ep) - \ep \ra_\Te^\text{PA}$ calculated from (\ref{eq:deN_PO_theory}) without the $-\pi/4$ factor and averaged over $\al$. Bottom: the same magenta line is shifted leftward by $115.6$. }\label{fig:deN_PA}
\end{figure}

We study SA and PA of the fluctuation of SS
\begin{eqnarray}
&\la \scr{N}(\ep) - \ep \ra^\text{SA}    \approx 0   \label{eq:EA_fluc_SA}\\
&\la \scr{N}(\ep) - \ep \ra^\text{PA}    \approx 0 . \label{eq:EA_fluc_PA}
\end{eqnarray}
In Figs. \ref{fig:deN_SA} and \ref{fig:deN_PA}, we present the results obtained with SA and PA respectively. For a large range of sampled energies, SA gives near zero result. PA produces regular oscillations about zero line. Theoretically, the oscillations are due to the sine term with PO-$(M,M)$ in (\ref{eq:deN_PO_theory}), which does not vanish upon PA. \cite{PA} In Fig. \ref{fig:deN_PA}, the theoretical result obtained from (\ref{eq:deN_PO_theory}) with PA needs to be shifted leftward to be consistent with the numerical result. This shift is due to the perimeter correction and can be calculated as $r=115.6$ from $\pi r^2/4 = 10500$, where 10500 is the average energy in Fig. \ref{fig:deN_PA}. We also observe that the factor $-\pi/4$ is critical for a good vertical fit.

The deviation of $\la \scr{N}(\ep) - \ep \ra$ from 0 in PA reveals a shortcoming of PA. But its small magnitude indicates PA is basically proficient as an EA method. An advantage of PA is that the distribution $f^\text{PA}(\al)$ works for any energy scale, while the range of sampled energies needs to grow as $\ep^{1/2}$ in SA.

\subsection{Interval level number variance}

\begin{figure}
\begin{center}
\includegraphics[width=0.45\textwidth]{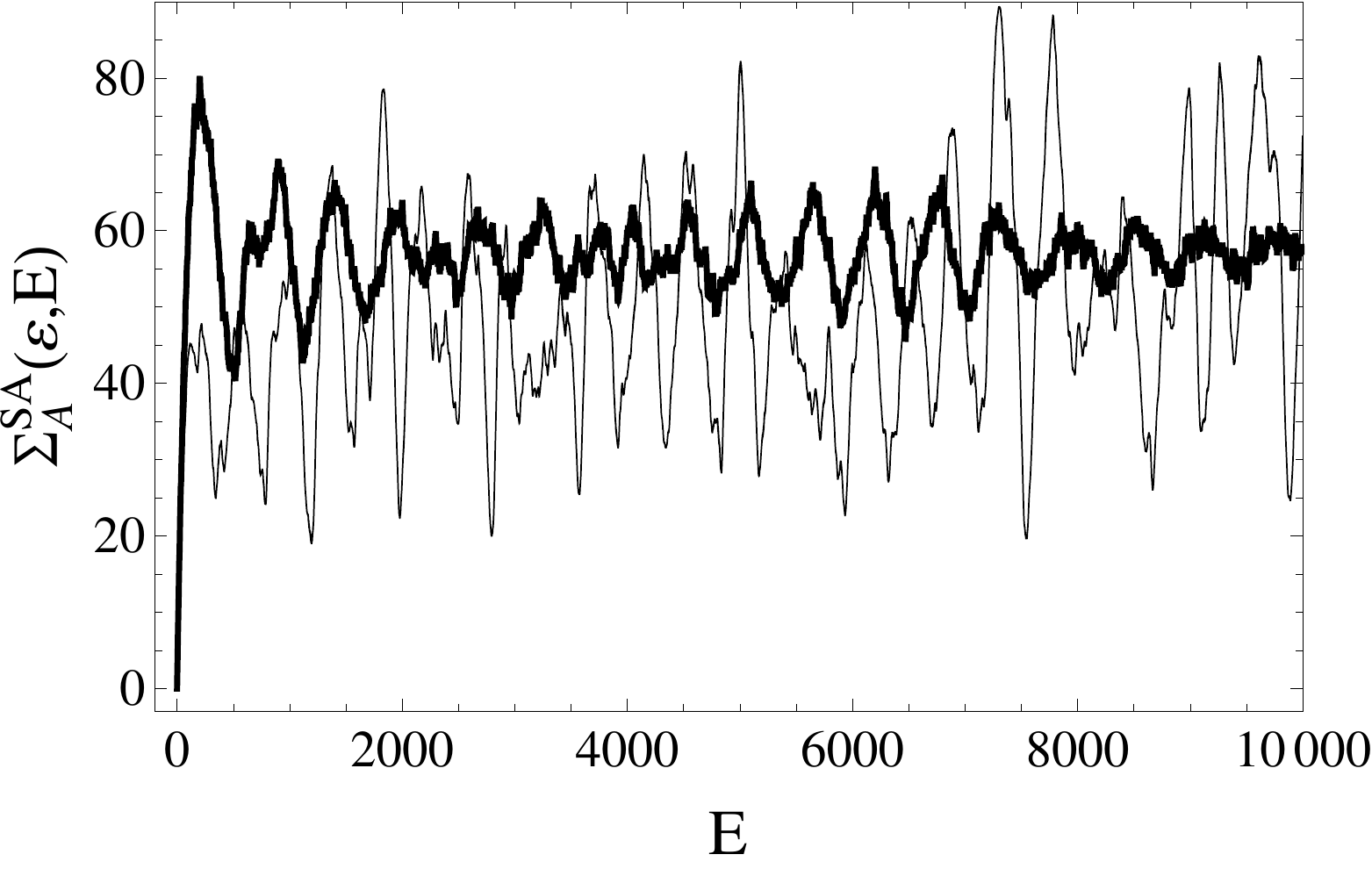}
\end{center}
\caption{IV calculated from SA. Thin and thick lines: the ranges of sampled energies are $[90500, 100500]$ and $[75000, 125000]$ respectively. }\label{fig:Si_SA}
\end{figure}

\begin{widetext}
\begin{figure}
\centering
\includegraphics[width=0.80\textwidth]{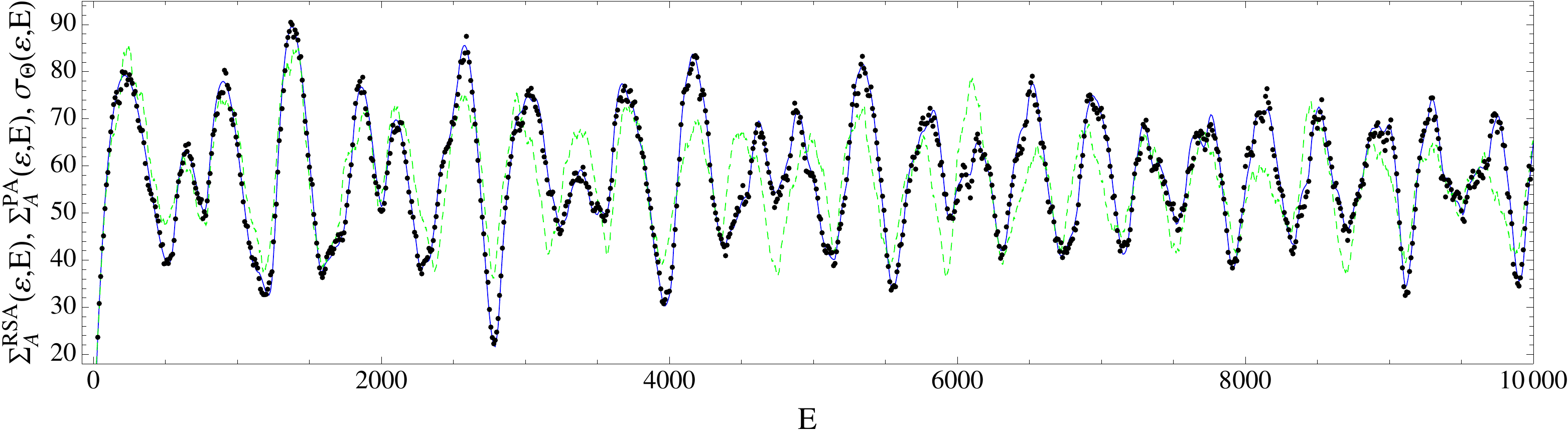}
\caption{IV calculated with RSA and PA. Black dots: RSA calculated from an ensemble of sampled energies in [$10^5$, $2\times 10^5$]. Green dashed line: PA. Blue solid line: theoretical result calculated from (\ref{eq:sample_IV_theory}). }\label{fig:Si_RSA_PA}
\end{figure}

In Fig. \ref{fig:Si_SA}, we present IV computed from SA. Clearly, SA cannot properly produce the persistent oscillations of IV. If the range of sampled energies is small, SA produces close to sample specific oscillations, indicating insufficient sampling. If the range of sampled energies is large, SA suppresses IV oscillations when the interval $E$ grows.

In Fig. \ref{fig:Si_RSA_PA}, we present IV computed from RSA and PA. We observe that RSA is in better agreement with the theoretical result, (\ref{eq:Si_RSA_theory}) (or, equivalently, (\ref{eq:IV_theory_PA_al0_def})) and \ref{eq:sample_IV_theory}, than PA.

\subsection{Correlation function of spectral staircase}

\begin{figure}
\centering
\includegraphics[width=0.75\textwidth]{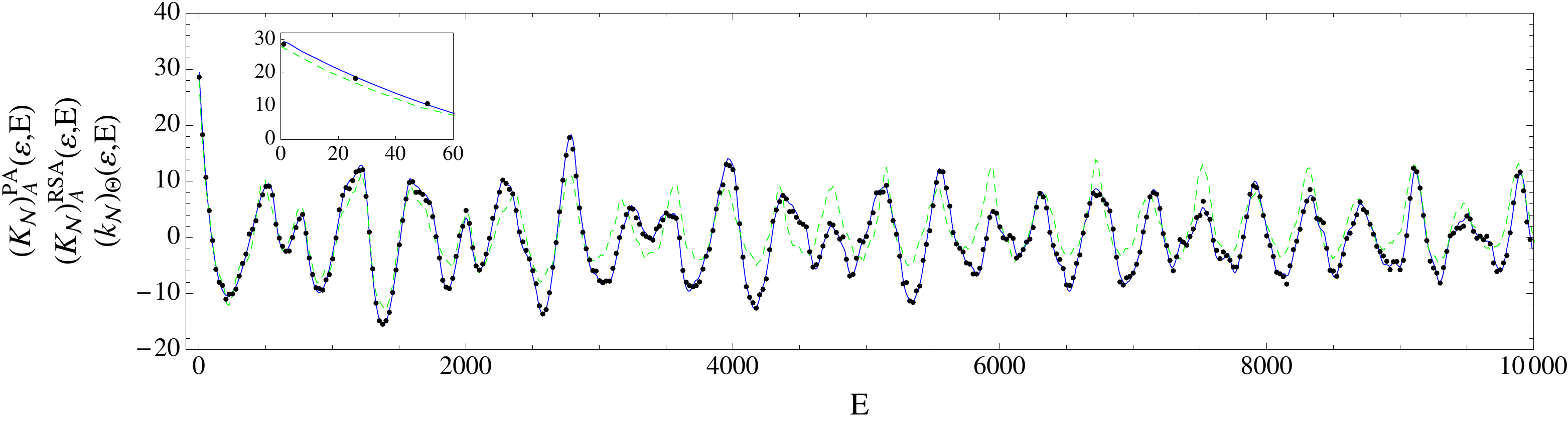}
\caption{CFSS with RSA and PA. Black dots: numerical result calculated from the definition of $K_\scr{N}$ in (\ref{eq:KN_def}) and averaged through RSA. Green dashed line: PA. Blue solid line: theoretical result calculated from the first Eq. (\ref{eq:relation_kN_de3_si}). Insert shows small $E$ behavior - close to linear, according to the second Eq. (\ref{eq:relation_kN_de3_si}). }\label{fig:KN_RSA_PA}
\end{figure}
\end{widetext}

In Fig. \ref{fig:KN_RSA_PA} we plot $K_\scr{N}(\ep,E)$ computed with RSA and PA. Again, we observe that RSA is in a better agreement with theoretical result than PA.

\subsection{Saturated spectral rigidity}

\begin{figure}
\centering
\includegraphics[width=0.45\textwidth]{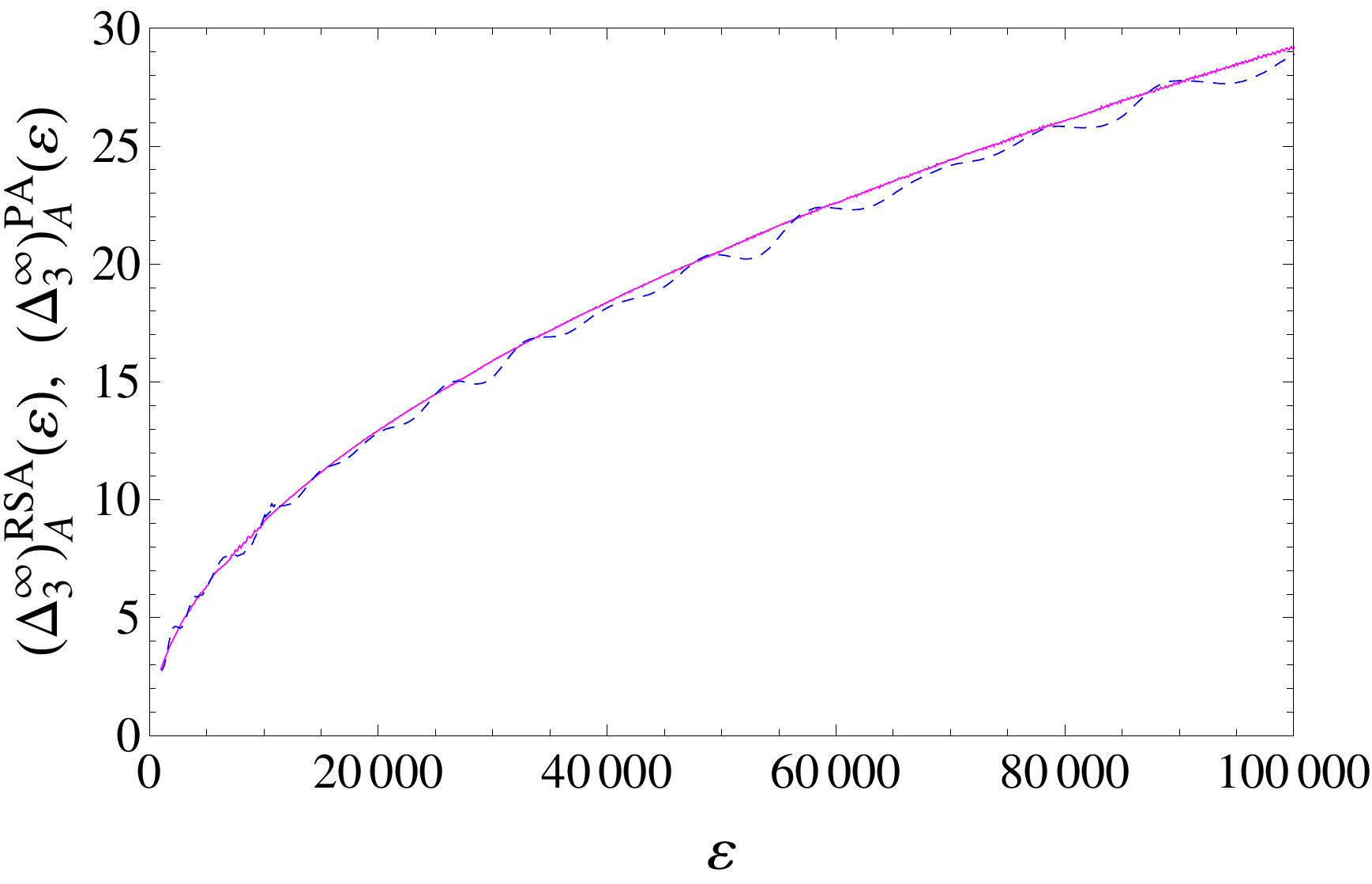}
\caption{Numerical result of saturation SR computed from RSA and PA. Magenta line: PA. Dashed blue line: RSA. For $(\De_3^\infty)_\text{A}^\text{RSA}(\ep)$, the range of sampled energies $[\ep, 2\ep]$. }\label{fig:Delta3_RSA_PA}
\end{figure}

\begin{figure}
\centering
\includegraphics[width=0.45\textwidth]{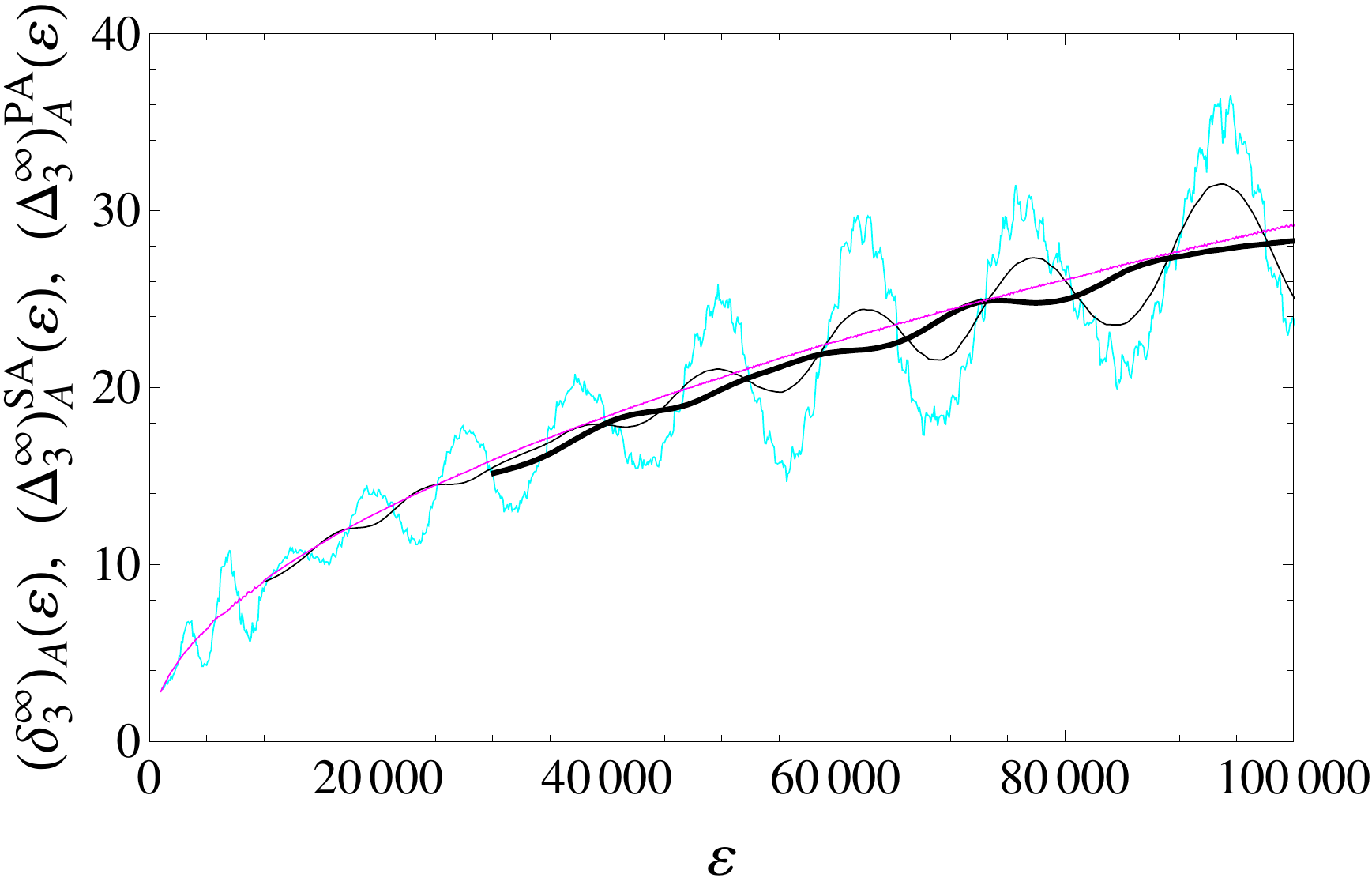}
\caption{Numerical results of saturation SR computed from SA and PA. Magenta line without any oscillations: PA result. Thin and thick black lines: SA with averaging range $10^4$ and $5 \times 10^4$ respectively. Jagged cyan line: sample saturation SR computed with $E=10^3$ for $\ep<=10^4$ and $E=5\times 10^3$ for $\ep>10^4$. }\label{fig:Delta3_SA_PA}
\end{figure}

In Fig. \ref{fig:Delta3_RSA_PA}, we present saturation SR computed from RSA and PA. Clearly, PA yields a better result since RSA shows small oscillations, while by theory (\ref{eq:sample_delta3_theory}) saturation SR should be a smooth function of $\ep$.

In Fig. \ref{fig:Delta3_SA_PA}, we present saturation SR computed with PA and SA and sample saturation SR (computed with (\ref{eq:sample_SR_def})). The latter shows large-range oscillations, which is absent in the PA result. If the range of sampled energies is sufficiently large, SA gives a result close to PA; otherwise, SA gives behaves similarly to sample specific SR.

\subsection{Global level number variance}\label{sec:num:GV}

The results of GV for four different integrable systems computed from PA are presented in \cite{ma12gv}. GV oscillates around saturation SR.
Unlike IV and CFSS, we can not find a rescaled form of of SA for GV. A simple definition of SA for GV is
\begin{equation}\label{eq:GV_as_De3_def}
\Si_g( \ep )
= \fr{1}{n+1} \sum_{i=0}^n [\scr{N}(\ep_i) - \ep_i ]^2 ,
\end{equation}
where the sampled energy $\ep_i$ is equally distributed in the range $[\ep-\om/2, \ep+\om/2]$. This is the definition of sample SR: $\de_3(\ep, \om)$. The integration in (\ref{eq:delta3_sample_GV}) (after we change $E$ into $\om$) can be approximated by the numerical integration as
$\fr{1}{n+1} \sum_{i=0}^n \si_g(\ep -\om/2 + i \om/n)$, which becomes (\ref{eq:GV_as_De3_def}) if $\ep_i = \ep -\om/2 + i \om/n$. We come to the conclusion that any SA is incapable of reproducing large oscillations of GV around SR. A detailed discussion of persistent oscillations of GV is given in \cite{ma12gv}.

\section{Conclusions}\label{sec:summary}

We introduced a new SA procedure -- RSA -- to cure some of the intrinsic problems of SA. 

For RB, we found that SA cannot produce persistent oscillations of IV and has some difficulties with SR. Any spectral averaging is unsuitable for GV oscillations. RSA is best suited for oscillations of IV and CFSS and generally works for SR, while PA is best suited for SR, GV and generally works for IV and CFSS.

Relative RSA success for SR in RB does not carry over to more complex system, such as Modified Kepler Problem \cite{ma10mk} and elliptic billiards \cite{ma11eb}, where SR exhibits non-trivial dependence on the running energy (spectral position) that RSA is incapable of yielding. 

To summarize our findings: PA always works numerically, RSA may be occasionally more accurate while traditional SA is almost always inadequate. We also have good agreement between theory and numerical results. The latter includes the fact that, with the exception of GV, DA yields sufficiently accurate predictions.

RSA should find its use in circular billiards, for which no proper PA procedure exists. On the other hand, PA may also find application to chaotic systems. For instance, PA of a Sinai billiard -- a circular hole in a rectangular billiard -- can be achieved through varying the aspect ratios of the sides of Sinai billiard.

\bibliographystyle{unsrt}

\begin{thebibliography}{1}
\bibitem{PA} Alternatively, one could use a different definition of theoretical PA, \cite{ma10mk,ma12gv} namely
\end{thebibliography}

\begin{eqnarray}
\Si_{\Te}^\text{PA} (\ep, E) &\equiv
\int \si_\Te (\ep, E) f^\text{PA}(\al) d\al , \label{eq:IV_theory_PA_def}
\end{eqnarray}
With this definition, an argument similar to Sec. \ref{subsec:linear_expansion_SA}, using (\ref{eq:sample_IV_theory}), shows that oscillations due to $(M_1,M_2)$ terms with $M_1\neq M_2$ would decay to the average value when $E$ becomes large. But for $(M,M)$ terms, this argument fails as the first order derivative over $\al$ vanishes (approximately) for $\al_0 \approx 1$ and the oscillations persist even when $E$ becomes large. Then the persistent oscillations of $\Si_\Te^\text{PA}(\ep, E)$ are solely due to PO-$(M,M)$. Replacing the sine terms in (\ref{eq:sample_IV_theory}) and (\ref{eq:IV_theory_PA_def}) with 1/2, except for $M_1= M_2$, we have for large $E$
\begin{equation}\label{eq:Si_RB_MM}
\begin{split}
&\Si_\Te^\text{PA} (\ep, E) \approx
4\sqrt{\fr{\ep}{\pi^5 }} \Bigg[ \sum_{M_1\neq M_2=0}^\infty \fr{\de_\mathbf{M}^2 }{2 R_\mathbf{M}^3} \\
& \quad\quad + \sum_{M_1 = M_2 = 1}^\infty  \fr{\int \sin^2\lf[\sqrt{\fr{\pi}{\ep} } R_\mathbf{M} E \rg] f^\text{PA}(\al) d\al}{R_\mathbf{M}^3} \Bigg].
\end{split}
\end{equation}
The implication of the above is that DA breaks down in such procedure. For IV, (\ref{eq:Si_RB_MM}) describes numerical results fairly well, even for large intervals. In order to obtain good agreement with numerical results for GV, however, the use of $ f^\text{PA}(\al)$ in theoretical averaging necessitates taking non-diagonal terms into account. For instance, the theoretical sample GV containing both diagonal and non-diagonal terms is given by \cite{ma12gv}
\begin{equation}\label{eq:sample_GV_theory}
\begin{split}
&(\si_g)_\Te(\ep) \\
&= \fr{4}{\hbar^{2\mu}} \lf[\sum_\mathbf{M} \fr{\de_\mathbf{M}
A_\mathbf{M}(\ep)}{T_\mathbf{M}(\ep)}
\sin\lf(\fr{S_\mathbf{M}(\ep)}{\hbar} -\fr{\pi}{4}\rg)\rg]^2 .
\end{split}
\end{equation}
One of the reasons to use theoretical $ f^\text{PA}(\al)$ averaging in the first place is to obtain agreement with numerical results for SS, as shown in Fig. \ref{fig:deN_PA}.

\end{document}